\documentstyle[multicol,prl,aps,epsf,psfig,epsfig]{revtex}

\begin{document}

\title{Free energy barrier for single-chain melting and crystallization}
  \author{Wenbing Hu$^1$, Daan Frenkel$^1$, Vincent B. F. Mathot$^2$\\$^1$FOM Institute for Atomic and Molecular Physics,\\ Kruislaan 407, 1098 SJ Amsterdam, The Netherlands \\$^2$DSM Research, P. O. Box 18 Geleen, The Netherlands}

\maketitle
\begin{abstract}
  In this paper, we report dynamic Monte Carlo simulations of melting and crystallization in a single-chain system. Their free energy barriers are calculated by the umbrella sampling method and can be described well by a simple expression $\Delta F = n\Delta f+\sigma (N-n)^{2/3}$, where $n$ is the amount of molten bonds, $\Delta f$ is the free energy change of each molten bond from a crystalline state, $N$ is the chain length, and $\sigma$ is the surface free energy of crystallite. We found that, together with the expression $\Delta F = n\Delta f+\sigma (N-n)^{1/2}$ for molecular nucleation, the molecular-weight dependent properties of the free-energy barriers for polymer primary and secondary nucleation, in particular, the molecular segregation during crystal growth, can be reproduced.  Then for the mechanism of polymer crystallization, we suggested a quantitative model of intramolecular nucleation, as a direct development from the previous qualitative description of molecular nucleation model.
\end{abstract}

\begin{multicols} {2}
\section{Introduction}  
     Polymer nucleation dominates the crystallization process in a quiescent liquid. \cite{1} The nucleation rate $G$ can be described as the formula proposed by Becker: \cite{2}
\begin{equation}
   G=G_0exp[(\Delta E+\Delta F_c)/(k_BT)],
\label{eq1}
\end{equation}
where $G_0$ is a prefactor, $\Delta E$ is the activation energy for diffusion across the phase boundary, $\Delta F_c$ is the free energy barrier for critical nucleus formation, and $k_B$ Boltzmann constant, $T$ temperature. After the primary nucleation, the secondary nucleation is believed to be a rate-determined step of polymer crystal growth in the bulk states, leading to a constant linear crystal growth rate and its proportionality with $exp(-1/T/\Delta T)$. \cite{3} For small molecules, the secondary nucleation is a two-dimensional nucleation on a smooth growth front as a key process to generate layer by layer advance.  The well-known Lauritzen-Hoffmann theory for polymer crystal growth was suggested based upon this kind of secondary nucleation process.\cite{4} However, the large-scale smooth front may not be available due to the thermodynamic roughening at high temperatures, and the ill conformation of chains on the growth front may construct an entropic barrier for polymer crystal growth rather than the surface free energy penalty.\cite{5} Moreover, the qualitative model of molecular nucleation was suggested to explain the molecular segregation with different chain length during polymer crystal growth.\cite{3} 

     Nevertheless, most of molecular-weight-dependent properties of polymer crystallization have not been explained well by these incompatible models.  Long chains require for large supercoolings in both primary nucleation \cite{6} and crystal growth than short chains, while the superheating in melting seems quite small and insensitive to the chain length.\cite{7} The linear crystal growth rate of polymers has been found to be significantly chain-length dependent.\cite{8,9} However recently, a master curve for the temperature dependence of crystal growth rate was found, which implies that the chain-length dependence should be attributed to the prefactor $G_0$ in Eqn. \ref{eq1} rather than the free-energy barrier $\Delta F_c$ for secondary nucleation.\cite{10} This chain-length independence was also observed in the primary nucleation rates of polyethylene.\cite{11} The complicated molecular-weight-dependent properties imply a different crystallization mechanism of polymers from that of small molecules. Therefore, understanding these properties is of fundamental importance.

   In principle, the molecular-weight-dependent properties can be attributed to the processes of diffusion or the free energy change, which are sensitive to the size of the whole molecules.  These kinds of process can be the intramolecular process since all the units of a molecule cannot crystallize simultaneously, especially for long flexible chains. When a polymer system performing melting and crystallization is limited to a single chain, the free energy change of intramolecular process becomes essential. A crystallite formed solely by a single chain has been observed in experiments \cite{12} and reproduced in many numerical simulations.\cite{13}  Recently, benefited from the simple lattice model \cite{14} and the umbrella sampling techniques \cite{15}, we can calculate the free energy change for phase transitions of a single-chain system through dynamic Monte Carlo simulations.\cite{16} In this paper, the chain-length dependence of the free energy barrier for phase transitions was investigated.

  The rest of this paper is organized as follows. After the simulation technique section, a simple estimation of the free energy change for a single-chain system was performed. Then, the results of estimation for the barrier of single-chain equilibrium states and of primary nucleation were compared to the simulation results and experiments, and the formula was extended to a secondary nucleation process. Most of important molecular-weight-dependent properties of polymer crystallization and melting can thus be discussed.  After that, we suggested an intramolecular nucleation model for the mechanism of polymer crystallization and discussed its advantages.

\section{Simulation techniques} 
   In our simulations, the lattice chain occupying a number of consecutive sites performs micro-relaxation in a cubic box with periodic boundary conditions.\cite{17} The size of box is $256 \times 256 \times 256$, which is large enough to avoid the contact from one side of the suspending single coil to the other side through the periodic boundaries.  The self- and mutual-avoiding jumps of the monomers on the lattice can be along either grid lines or diagonals, so the coordination number of cubic lattice is $9+8+9=26$. Only the parallel packing tendency $E_p$ for each pair of the non-bonded bonds in neighbor is employed as the driving force for crystallization, while the bending penalty along the chain and the mixing heat of the chain units in a solvent are set to zero for simplicity.\cite{14}  The system temperature is then represented as a dimensionless term $T^*=k_BT/E_p$ in the Metropolis sampling. 

     A series of single-chain systems containing variable chain units (16, 32, 64, 128, 256, 512, 1024) were investigated. The sample, prepared initially in a regularly folded state, relaxed down to the equilibrium state at the preset temperature.  The degree of melting is traced as the number of molten units, which is defined as the bond containing less then five bonds packing around in parallel.  The free energy distribution to the number of molten units was estimated according to the Boltzmann relation
\begin{equation}
   \Delta F/(k_BT)=-ln P(n),
\label{eq2}
\end{equation}
where $P(n)$ is the probability to find a sample having $n$ molten units. The umbrella sampling method permits us to calculate the free energy distribution in a window away from the most favorable state in equilibrium.\cite{18} A parabolic biasing potential was employed here.\cite{19} The results for free energy distribution in several such kind of windows covering over a wide range of $n$ were then merged together through parallel tempering. \cite{20}

\section{Theoretical estimation}
   The free energy change of a single-chain system can be estimated with a fully ordered ground state in the bulk.  In the simple lattice model, the potential energy loss can be estimated as $\Delta E=nE_p(q-2)/2$, where $n$ is the number of molten units, $q$ is the coordination number of a regular lattice, the first $2$ is the number of connective bonds along the chain, so $q-2$ is the possible number of the parallel packing bonds around each bond, and the second $2$ is a symmetrical factor for bond-bond pair interactions.  The interfacial free energy penalty is estimated to be $\sigma(N-n)^{2/3}$, where N is the total number of bonds on the chain, and $\sigma$ is the surface free energy.  The entropy gain should be attributed to the conformational entropy change of the molten bonds, and as a first approximation, their conformation can be in analogy with a self-avoiding walk.\cite{21} Therefore,  $\Delta S=n \text{ln} \mu + (\gamma-1) \text{ln} n$.  Here, $\mu$ is the connective constant, which is the number of possible directions for each step of random walk, and for chain conformation, $\mu=q-1$;  $\gamma$ is a critical exponent.  The term $\text{ln}n$ is a correction term for the self-avoiding walks deviated from the random walk and be actually equivalent to the mixing entropy of a single coil.  This nonlinear $\text{ln}n$ term is expected to be much smaller than that linear $n$ term when $n$ is large. Although the molten units may belong to several segments of the chain tied on the surface of crystallite, their linear terms contributing to the conformational entropy are additive. Therefore, we neglected the nonlinear term for simplicity.  The free energy change of a single-chain system can be estimated as 
\begin{eqnarray}
  \Delta F = \frac{q-2}{2} n E_p + \sigma (N-n)^{2/3} - k_BT n \text{ln}(q-1) \nonumber \\  =n\Delta f + \sigma (N-n)^{2/3}, 
\label{eq3}
\end{eqnarray}
where $\Delta f=E_p(q-2)/2- k_BT \text{ln}(q-1)$ is the body free energy change of each molten unit. Corresponding to our simulation model, $N$ is fixed, $q=26$, only $\sigma$ and $T$ are required to be determined.

  The formula of Eqn. \ref{fig3} reflects the Gibbs phenomenological expression for nucleation of phase transitions.   At the critical size (top of free energy curve), the free energy change can be calculated from $\partial{\Delta F}/ \partial{n}=0$ as $N \Delta f+ 4 \sigma^3 / (27\Delta f_{eq}^2 )$.  At the fully crystalline state, $n=0$ and hence the free energy change is $\sigma N^{2/3}$, while at the coil state, $n=N$ and hence the free energy change becomes $N\Delta f$.  Therefore, the height of free energy barrier can be calculated from the difference of free energy changes between the initial and the critical states.

\begin{figure}[h]
\centering\epsfig{file=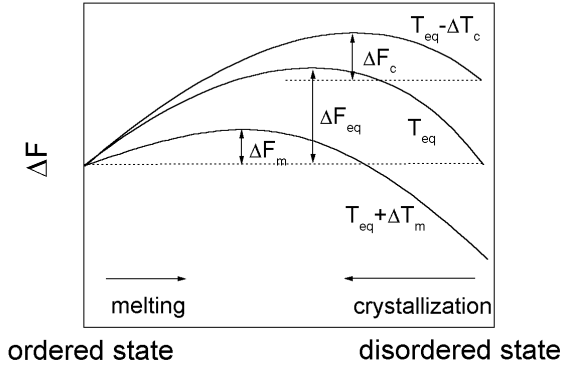,
         height=6 cm}
\caption{Schematic showing free energy curves in the vicinity of equilibrium melting temperature $T_{eq}$, the critical free energy barrier $\Delta F_{eq}$, and the free energy barriers for melting $\Delta F_m$ with a superheating $\Delta T_m$ and for crystallization $\Delta F_c$ with a supercooling $\Delta T_c$.
}
\label{fig1}
\end{figure}

  Before starting the next section, several terms should be clarified as schematically shown in Fig. \ref{fig1}.  When the lowest point of free energy change at the left-hand side (the crystalline state) has the same level as that at the right-hand side (the melt state), the sample system is in equilibrium between melting and crystallization. This temperature is the equilibrium melting point $T_{eq}$. The free energy barrier at this temperature is the equilibrium free energy barrier $\Delta F_{eq}$. At a slightly high temperature $T$, the crystalline state becomes metastable, and the free energy barrier for melting is $\Delta F_m$ with a superheating $\Delta T_m=T-T_{eq}$; at a slightly low temperature $T$, the melt state becomes metastable, and the free energy barrier for crystallization is $\Delta F_c$ with a supercooling $\Delta T_c=T_{eq}-T$.

\section{Results and discussion}
\textbf{1. Single-chain primary nucleation}

   Figure 2 shows the free energy changes for phase transitions of a single 1024-mer, starting from the initial crystalline state located at zero-line level. The equilibrium melting point was found to be $2.967 E_p/k_B$. Deviating from this temperature, there will be a supercooling or a superheating to drive crystallization or melting respectively. 

\begin{figure}[h]
\centering\epsfig{file=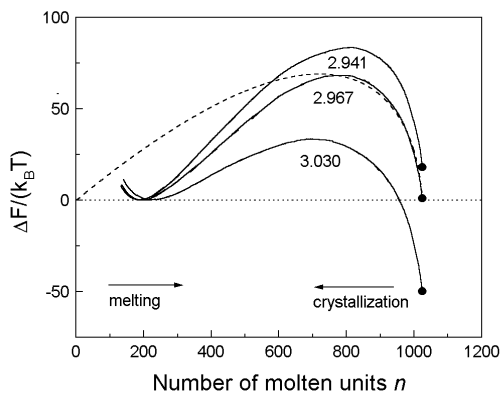,
         height=6 cm}
\caption{Free energy curves vs. the number of molten units for 1024-mer at the denoted temperatures with the unit of $E_p/k_B$.  The numerical results (solid lines) are calculated by umbrella sampling with 15 windows.  The dashed line is calculated from Eqn. \ref{eq3} with $\sigma=15E_p$ and the fitted equilibrium melting temperature $3.2657 E_p/k_B$.
}
\label{fig2}
\end{figure}

  The height of equilibrium free energy barrier in simulations can be fitted by the simple estimation of free energy change in Eqn. \ref{eq3}. As demonstrated by the dashed curve in Fig. \ref{fig2}, a proper set of $\sigma$ and $T$ can fit the height of free energy barrier well. The significant deviation at the small amount of molten units should be attributed to some molten units of simulations located at the fold-end surface of the crystallite and the nonlinear entropy contribution in theoretical estimation neglected for simplicity.  We are not going to improve the precision of estimation at small molten units, since here we only focus our attention on the height of free energy barrier for phase transitions.
  
\begin{figure}[h]
\centering\epsfig{file=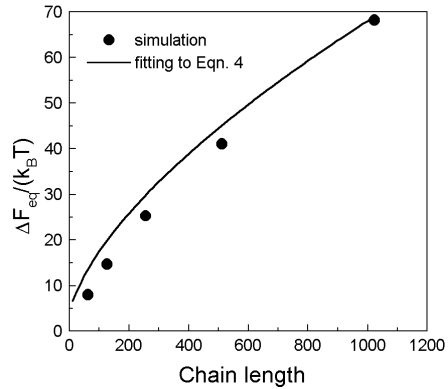,
         height=6 cm}
\caption{Height of equilibrium free energy barrier vs. chain length.  The solid curve is calculated from Eqn. \ref{eq4} with $\sigma=15E_p$.
}
\label{fig3}
\end{figure}

   At the equilibrium melting temperature, the coexistence of the crystalline and amorphous states gives $\sigma N^{2/3} =N\Delta f_{eq}$, or $\Delta f_{eq}=\sigma N^{-1/3}$.  The height of equilibrium free energy barrier can be estimated as 
\begin{equation}
\Delta F_{eq}=\frac{4 \sigma^3}{27 \Delta f_{eq}^2}=\frac{4}{27} \sigma N^{2/3}.
\label{eq4} 
\end{equation}
So the height of equilibrium free energy barrier is expected to increase monotonically with the chain length. This is exactly what we observed in Fig. \ref{fig3}, where the simulation results for the height of equilibrium free energy barrier with variable chain length are quite close to the fitting curve of Eqn. \ref{eq4} with just a single set of parameter $\sigma=15E_p$.

\begin{figure}[h]
\centering\epsfig{file=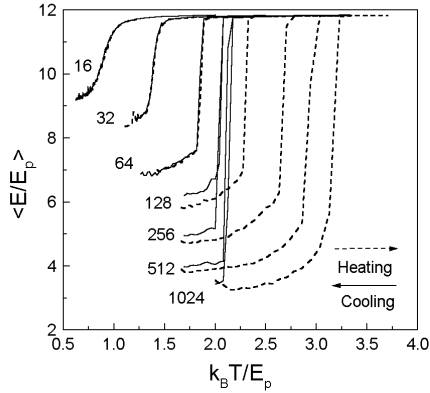,
         height=6 cm}
\caption{Heating and cooling curves (dashed and solid lines respectively) of the potential energy with the denoted variable chain length (number of units). The potential energy is defined as the mean value of $24-p$, where p is the number of parallel bonds packing around each bond. The heating and cooling programs are the steps in the value of $E_p/(k_BT)$, each having a length of $0.01$ and a period of $10^6$ Monte Carlo (MC) cycles. One MC cycle is defined as one jump for each chain-unit on average. The reported data are averaged over $1000$ samples, each with $1000$ MC-cycle interval.  }
\label{fig4}
\end{figure}

\begin{figure}[h]
\centering\epsfig{file=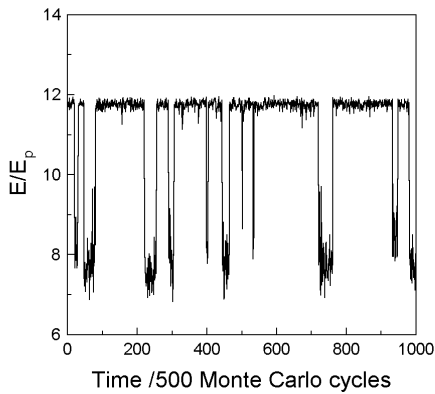,
         height=6 cm}
\caption{Isothermal curve of the potential energy for 64-mer at $T=1.845 E_p/k_B$. The initial state was picked up from the corresponding heating curve in Fig \ref{fig4} after $10^6$ MC-cycle annealing at this temperature step.
}
\label{fig5}
\end{figure}

   The monotonic chain-length dependence of equilibrium free energy barrier implies that if the chain length is small enough, the height of this barrier can be smaller than the thermal fluctuation level. In other words, short chains should show no hysteresis in the temperature scanning for melting and crystallization.  This can be identified in our simulations as demonstrated in Fig. \ref{fig4}, where in short-chain cases, the phase transitions are reversible and appears continuous with temperature changes.  Actually, this continuous change should be attributed to the average of packing potential energy in each step of temperature change.  If we trace an isothermal process in the vicinity of phase transition temperature of 64-mer, as shown in Fig. \ref{fig5}, the single-chain system is quickly jumping between two extreme states and shows all-or-none feature. So during phase transition of short chains, the temperature change just shifts the relative probability of two states from one to another and leads to a continuous change in the average of packing potential energy.

      When the temperature leaves away from the equilibrium point, the height of free energy barrier for melting can be estimated as
\begin{equation}
\Delta F_m=\Delta f N-\sigma N^{2/3}+\frac{4\sigma^3}{27\Delta f^2},
\label{eq5} 
\end{equation}
meanwhile, the height of free energy barrier for crystallization is
\begin{equation}
\Delta F_c=\frac{4\sigma^3}{27 \Delta f^2}.
\label{eq6} 
\end{equation}
At a fixed temperature (and hence $\Delta f$), $\Delta F_m$ depends upon the chain length, but $\Delta F_c$ seems not.
   
\begin{figure}[h]
\centering\epsfig{file=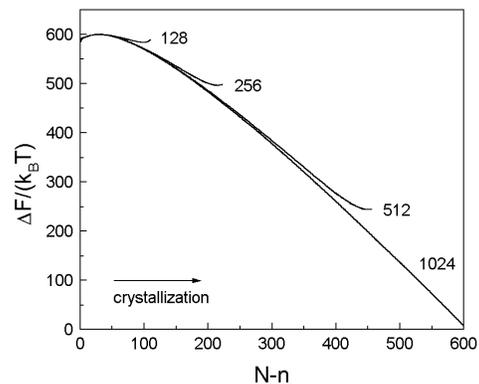,
         height=6 cm}
\caption{The free energy changes of single-chain systems with the denoted variable chain length (number of units) at the same temperature $T=2.174 E_p/k_B$. Curves are shifted to meet at their tops. 
}
\label{fig6}
\end{figure}

      The chain-length independence of the free energy barrier for primary nucleation is in consistence with recent experimental reports about polyethylene primary nucleation rates.\cite{11}  This reveals the importance of intramolecular process in polymer primary nucleation.  Actually, this independence is also verified in Fig. \ref{fig4}, where the cooling curves show the crystallization temperatures, which are insensitive to the chain length when the chain length is large enough to show the hysteresis.  In the free energy calculation of simulations, we fixed the crystallization temperature there and found that the height of free energy barriers for crystallization is really insensitive to the chain length, as demonstrated in Fig. \ref{fig6}.

\begin{figure}[h]
\centering\textbf{a.}\epsfig{file=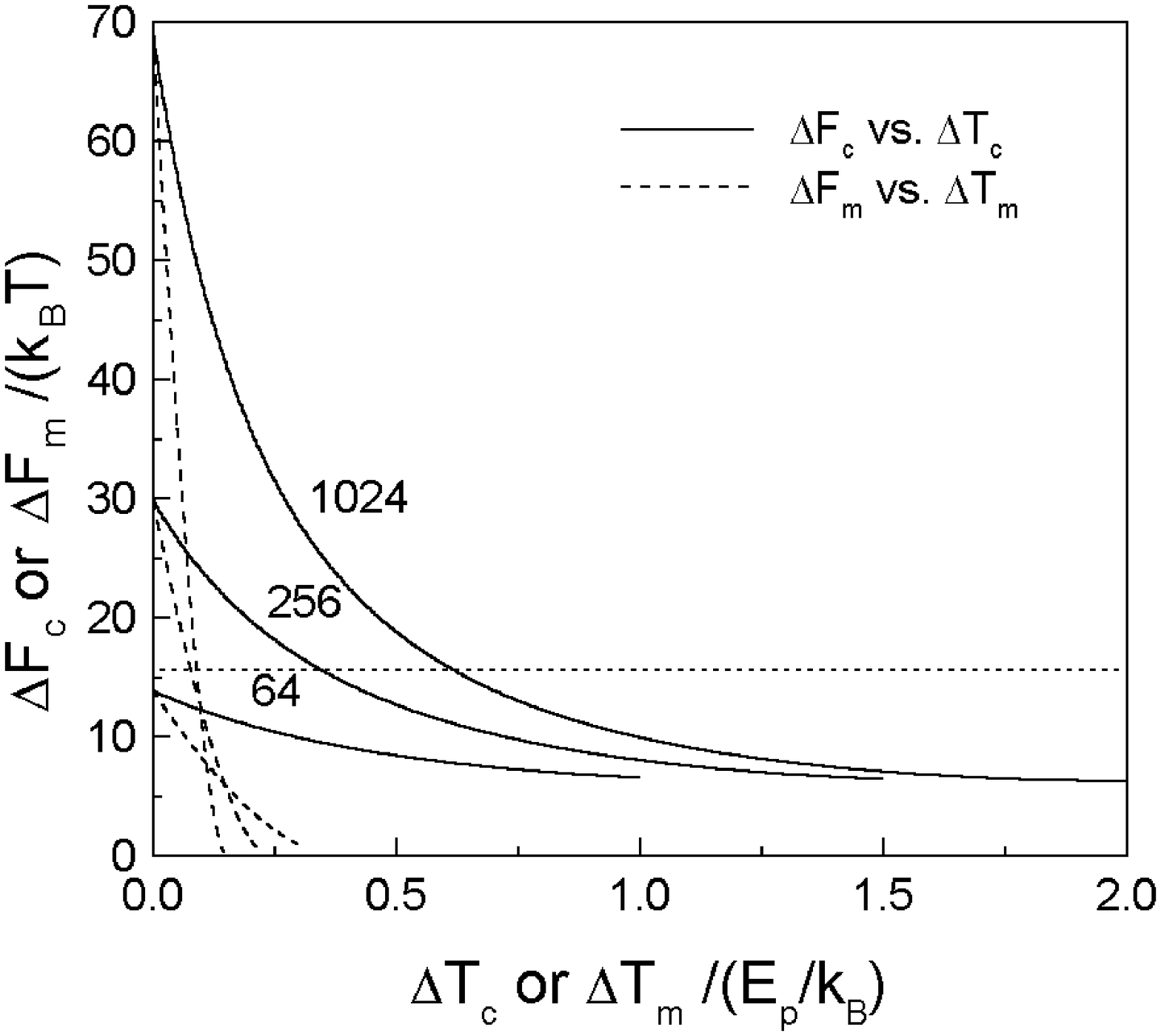,
         height=6 cm}
\textbf{b.}\epsfig{file=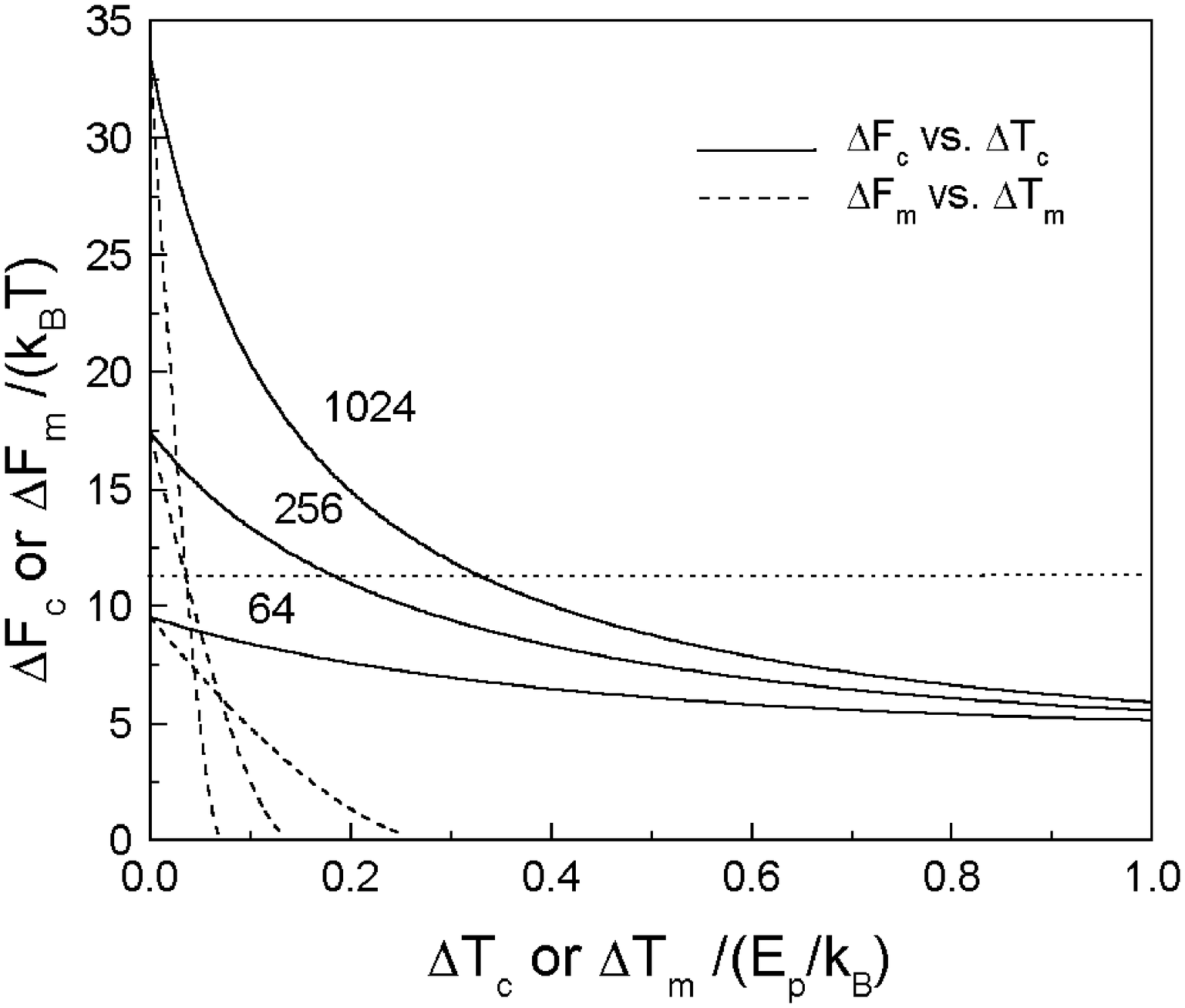,
         height=6 cm}
\caption{Height of free energy barriers for primary nucleation or melting (a) and secondary nucleation or melting (b) ($\Delta F_c$ vs. $\Delta T_c$ (solid lines) and $\Delta F_m$ vs. $\Delta T_m$ (dashed lines) with denoted chain length (number of units), calculated from Eqn. \ref{eq5} and \ref{eq6}(a), and Eq. \ref{eq8} and \ref{eq9} (b) with $\sigma=15E_p$.  The dotted line is the supposed thermal fluctuation level.
}
\label{fig7}
\end{figure}

    The supercooling or the superheating required for phase transitions are the temperatures at which the thermal fluctuation can get over the free energy barrier for phase transitions and generate crystallization or melting spontaneously.  The chain-length independence of $\Delta F_c$ at the crystallization temperatures does not mean the same independence at the fixed supercoolings, since the equilibrium melting point, i.e. the reference temperature for supercoolings, may be sensitive to the chain length.  Figure \ref{fig7}a shows the estimated free energy barriers for crystallization vs supercoolings, as well as that for melting vs superheatings according to Eqn. \ref{eq5} and \ref{eq6}.  At a supposed thermal fluctuation level shown in the dotted curves, the height of free energy barrier for crystallization is much more sensitive to the chain length than that for melting.  This is also consistent with the experiments for the bulk polymer systems, which show that long chains require large supercoolings for primary nucleation, while the superheating required for melting seems insensitive to the chain length.\cite{6,7} Therefore, the rising of supercooling for bulk primary nucleation of long n-alkanes can be associated to their intramolecular chain-folding process. 

\textbf{2. Extension to secondary nucleation}
 
  The success of the simple free energy estimation for polymer primary nucleation encourages us to extend our free energy estimation to secondary nucleation.  Following the idea of molecular nucleation, the first part of molecule growing onto a smooth front will incur a penalty for a two-dimensional nucleation. So corresponding to Eqn. \ref{eq3}, the free energy change of a single-chain growth onto a smooth front is
\begin{equation}
      \Delta F =n\Delta f + \sigma (N-n)^{1/2}. 
\label{eq7}
\end{equation}   
And correspondingly, the height of free energy barrier for melting can be estimated as
\begin{equation}
\Delta F_m=N\Delta f-\sigma N^{1/2}+\frac{\sigma^2}{4\Delta f},
\label{eq8} 
\end{equation}
and the height of free energy barrier for crystal growth is
\begin{equation}
\Delta F_c=\frac{\sigma^2}{4 \Delta f}.
\label{eq9} 
\end{equation}
Again, at a fixed temperature (and hence $\Delta f$), $\Delta F_m$ depends upon the chain length, but $\Delta F_c$ seems not. The chain-length independence of free-energy barrier for crystal growth agrees well with the experiments as mentioned master curves in the introduction section.\cite{10} Similar to primary nucleation and melting, the chain-length dependence of supercooling and superheating required to generate crystal growth and melting, see Fig. \ref{fig7}b, is also in consistence with experiments. \cite{7}

\begin{figure}[h]
\centering\epsfig{file=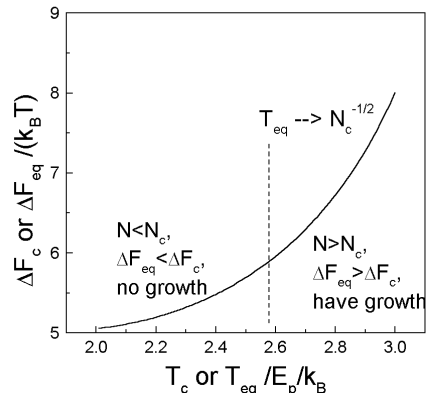,
         height=6 cm}
\caption{Height of free energy barriers for crystal growth as a master function of temperatures estimated from Eqn. \ref{eq9} and Eqn. \ref{eq10}. The dashed line is indicating a fixed crystallization temperature for discussion.}
\label{fig8}
\end{figure}

\begin{figure}[h]
\centering\epsfig{file=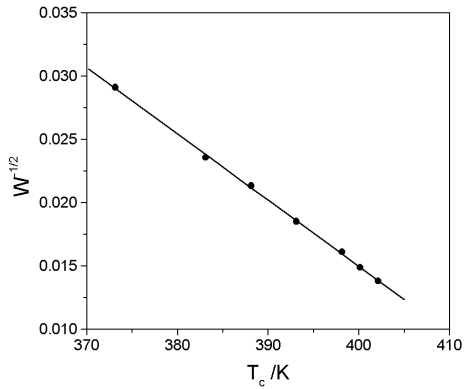, height=6 cm}
\caption{Equilibrium critical molecular weight $W^{-1/2}$ of polyethylene melt as a function of crystallization temperature $T_c$.  The data points are from Ref. 3 page 101 in Table V-7, Type M series, excluding the low molecular-weight end.  The fitting line shows a correlation coefficient of 0.9997. 
}
\label{fig9}
\end{figure}

  The equilibrium barrier $\Delta F_eq$ can also be obtained with the same formula as $\Delta F_c$, 
\begin{equation}
\Delta F_{eq}=\frac{\sigma^2}{4 \Delta f_{eq}}.
\label{eq10} 
\end{equation}
which implies a master curve for their temperature dependence. As demonstrated in Fig. \ref{fig8}, if the sample system is fixed at a temperature indicated by a dashed vertical line, corresponding to an equilibrium melting point at this temperature, there is a critical chain-length $N_c$ for a polymer mixture with wide chain-length distributions. For the polymers with chain length larger than $N_c$, they have the same height of free energy barrier for crystal growth at this temperature (see Eqn. \ref{eq9}), while their barrier for melting is much higher. This means that they can grow onto the front with enough stability. On the other hand, for those polymers with chain length smaller than $N_c$, they cannot stay on the growth front because of their lower barrier for melting than that for crystallization.  So, $N_c$ is a critical value for molecular fractionation during polymer crystal growth. According to the equilibrium between two states at this temperature, $\Delta f_{eq} N=\sigma N^{1/2}$, we obtain that $T_{eq}~N^{-1/2}$. This linear relationship can be well verified by the plot of experimental data for the equilibrium critical molecular weight of polyethylene melt fractionation at a crystallization temperature, see Fig. \ref{fig9}.

   Figure \ref{fig8} also demonstrates the molecular segregation in the crystallites grown at different stages of a cooling process. When the sample system is cooled from high temperatures, the critical chain length is shifting to smaller.  In a polymer mixture with a wide chain-length distribution, the longest chains will first approach the critical chain length on cooling, and therefore, contribute to the thick dominant lamellae, whereas those short chains approach the critical chain-length at the late stage of cooling and then contribute to the thin subsidiary lamellae.

\textbf{3. Intramolecular nucleation model}

  Our estimation about free energy change of single-chain melting and crystallization gives a perfect quantitative description to the molecular nucleation model. However, the original molecular nucleation model suggested only for the first part of molecule that incurs a free energy barrier for crystal growth onto the front.  Actually in polymer crystallization, after the first part of molecule has finished the nucleation process, the fast growth of the rest part may quite easily be stopped, such as by the entanglement with other chains or by the impingement of the two-dimensional nucleus with other nuclei on the front surface. The entanglement has been found to make long-term pauses during single-chain crystal growth in the semi-dilute solution. \cite{22} The impingement can be associated to the granular texture on the growth front observed recently under atomic force microscopy (AFM).\cite{23}  

\begin{figure}[h]
\centering\epsfig{file=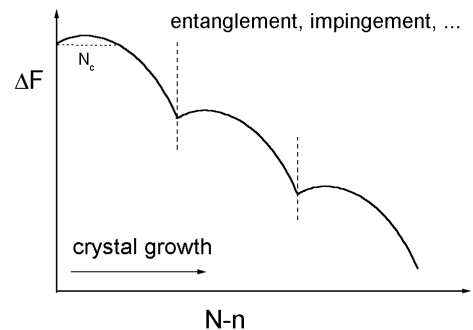,
         height=6 cm}
\caption{Schematic of the multiple nucleation processes of single-chain crystal growth. The dashed vertical lines are indicating the stops of crystal growth due to some kinds of reason such as chain entanglement in the melt or nuclei impingement on the front.}
\label{fig10}
\end{figure}

   Therefore, the rest part of chain may still need to perform secondary nucleation process for further crystal growth. If this rest part of chain is longer than the critical chain length, as schematically shown in Fig. \ref{fig10}, the free energy barrier for secondary nucleation, which is independent of the rest chain length, can still be got over in subsequent crystal growth.  Since the coil size of the long-chain molecule can be much larger than the thickness of the lamellar crystallite, the multiple nucleation process may not be necessary to happen one-by-one in sequence like shown in Fig. \ref{fig10}, rather, they can occur simultaneously when the nuclei are in enough distance on the same front or belong to different crystallite fronts.  

     Based upon the above considerations, we suggest an intramolecular nucleation model.  According to this model, any part of a molecule, if it is longer than the critical chain-length, can perform secondary nucleation process.  If the rest part of chain is shorter than the critical chain-length, it will become a cilium tied on the fold surface or lateral surface of the lamellar crystallites. This multiple nucleation model can reproduce a reasonable trajectory of a single chain in a semi-crystalline contexture, where one molecule may pass through several lamellae.

\section{conclusion}

   Simple free energy estimation for the barriers of single-chain melting and crystallization can fit well to our simulation calculations and be consistent with the experimental observations about their chain-length dependence. Then, most of important molecular-weight dependent properties about polymer melting and crystallization become rational under the consideration of intramolecular nucleation model. This proves the importance of intramolecular process in both primary nucleation and secondary nucleation processes of polymers. The advantages of intramolecular nucleation model, which directly develops from the idea of molecular nucleation model, include 1. reproduce the temperature dependence of nucleation rate as $lnG~1/T/ \Delta T^2$ for primary nucleation and $lnG ~ 1/T/ \Delta T$ for secondary nucleation;
2. in a comparison with Lauritzen-Hoffman model, avoid the assumptions such as the barrier of the first stem and a large smooth front for nucleation of new layer-growth;
3. in a comparison with Sadler-Gilmer model, the detailed process of crystal growth is much clear;
4. reproduce molecular segregation during polymer crystal growth;
5. reproduce the chain-length dependence phenomena of supercooling required for primary nucleation and crystal growth or superheating required for melting;
6. reproduce the chain-length independence of free energy barrier for both primary nucleation and secondary nucleation;
7. the multiple nucleation process can generate a reasonable trajectory of a molecule in the semi-crystalline contexture;
8. the impingement of the intramolecular nuclei on the front may lead to a granular texture;
9. reveal the importance of chain entanglement on the crystal growth rate.

  The diffusion process may contribute to the molecular-weight dependence of nucleation and crystal growth rate. It includes both the chain motion in the crystalline region and in the melt region that beyond the scope of our current discussion for free energy changes.  The lamellar thickness is a result of late stages of crystal growth, so we did not discuss here. The intramolecular nucleation model does not reject the intermolecular nucleation process, which may be dominant for the crystallizaiton from the dilute solution, especially for those short chains.  We expect to get much more insights about the mechanism of polymer crystallizaiton from the intramolecular nucleation model in near future.

\textbf{Acknowledgement.} W.H. thanks Dr. Bernhard Wunderlich for his stimulating discussions.  This work was financially supported by DSM Company and the division of Chemical Science of the Netherlands Organization for Scientific Research (NWO).  The work of the FOM Institute is part of the research program of FOM and is made possible by financial support from the NWO.


\end{multicols}

\end{document}